\documentclass[preprint,1p,12pt]{elsarticle}
\usepackage{epsfig}
\usepackage{hyperref}
\usepackage{multirow} 
\begin{document}
\begin{frontmatter}
 \title{Futures market efficiency diagnostics via temporal two-point correlations. Russian market case study.}
 
 \author[mp]{M.L.~Kopytin\corref{cor1}}
 \ead{m.kopytin@maragingpartners.ru}

 \author[mp]{E.V.~Kazantsev}
 \ead{e.kazantsev@maragingpartners.ru}

 \cortext[cor1]{Corresponding author}
 
 \address[mp]{Maraging Partners LLC, Moscow, Russia \\ \href{http://maragingpartners.ru}{http://maragingpartners.ru}}


\begin{abstract}
Using a  two-point correlation technique, we study emergence of market efficiency in the emergent Russian futures market by focusing on lagged correlations. The correlation strength of leader-follower effects in the lagged inter-market correlations on the hourly time frame is seen to be significant
initially (2009-2011) but gradually goes down, as the erstwhile leader instruments -- crude oil, the USD/RUB exchange rate, and the Russian stock market index -- seem to lose the leader status. 
 An inefficiency index, based on two-point correlations, is proposed and its history is established.
\end{abstract}
\end{frontmatter}

\section{Introduction}
\label{intro}
Success of algorithmic trading has as much to do with the market as with the nature of the algorithm. 
While the field of adaptive algorithms has its own problems and methods related to machine learning, 
 the phenomenology of just what the machines are ``learning'' can manifest itself and be captured
 in the N-dimensional  space enclosing the sets of observables (N-tuples) originating in the time series. 
The simplest non-trivial case of such spaces is the two-dimensional space
of pairs of quantities, where the object of learning would be the two-point correlation function (lossy information-wise in a non-stationary problem).
We use discrete time series and this function is defined on a discrete set of time lag values.
Zero values of this function at non-zero time lags would justify calling the time series a martingale\cite{martingale}, failing to establish
a deviation of the market from the efficient market hypothesis (EMH). Curiously, absence of features at non-zero time lags leads to maximum possible symmetry of the pair space, allowing one to think of the EMH as a form of symmetry principle.

Financial regulators and public policy makers have to understand the effect of futures speculation on prices and volatility, especially with regard to quantitative trading.
One has to venture outside the EMH to address this question.
While successful speculation\footnote{We focus on futures markets where the term ``speculator'' has no negative connotation, unlike the stock market where one can discuss ``intelligent investors'' as opposed to speculators. To us, speculators are futures market participants with no commercial interests in the commodities being traded.} enhances market efficiency on a certain time scale, it may do so at the cost of changing the  
market volatility on the same or different time scales. What happens with volatility when the inefficiency\footnote{In our language, the positive concept denoted by {\em market inefficiency} -- describing an information-rich state, the opposite of a trivial state of maximum symmetry -- is made to look negative, for inefficiency is a negation of {\em efficiency}. We will continue using these terms for lack of better ones.} is reduced, everything else being equal,
depends on the shape of the correlation function and that includes non-zero time lags. 

Non-commercial market participants, in particular, quantitative hedge funds, their marketers, clients and their advisors,
lack standard yardsticks characterizing the level of market efficiency within certain geographical and thematic domains, on certain time scales. 
Two-point autocorrelations and intermarket correlations, including non-zero time lags, suggest a solution which is transparent and expandable. 
Our case study, naturally, is focused on the instruments available within our region.
We see this as the first step towards establishing a set of such benchmarks aiding in rational evaluation
of manager performance and informing selection of regions, markets and strategies.

FORTS\footnote{An abbreviation for Futures and Options on RTS, Eastern Europe's largest derivatives exchange, based in Moscow.}, the Russian futures market,
 presents an interesting study case where, due to the emerging nature of the market, one can witness the emergence of market efficiency with contemporary electronic data, and even take part in exploiting the residual inefficiencies. 
We believe that the insights gained thereby --- the subject of this article --- are of general interest.

\section{Data and methods}
\label{sec:data}

We have analyzed data on the following five of the FORTS futures contracts, 
representing independent factors of macro-economic significance and offering highest liquidity.
The following abbreviations are used:
\begin{itemize}
\item RI: the nearest three-month contract on Russia's RTS stock market index
\item BR: the nearest one-month contract on Brent oil, representing a proxy for one of Russia's leading export commodities, crude oil
\item ED: the nearest three-month contract on EUR/USD
\item EU: the nearest three-month contract on EUR/RUB, an important hedging instrument for Russia's importers
\item SI: the nearest three-month contract on USD/RUB, an important hedging instrument for Russia's commodity exporters
\end{itemize}

The data are provided publicly by the FORTS exchange itself. FORTS is a modern electronic market of an auction type where all participants, including retail ones, enjoy direct market access via network. The original data are tick data which we aggregate to form hourly and daily bars. Both original and aggregated data are stored in MySQL databases. A custom API has been designed to interface our C++ analysis code with the databases.

\begin{figure}[t]
\epsfxsize=13cm
\epsfbox{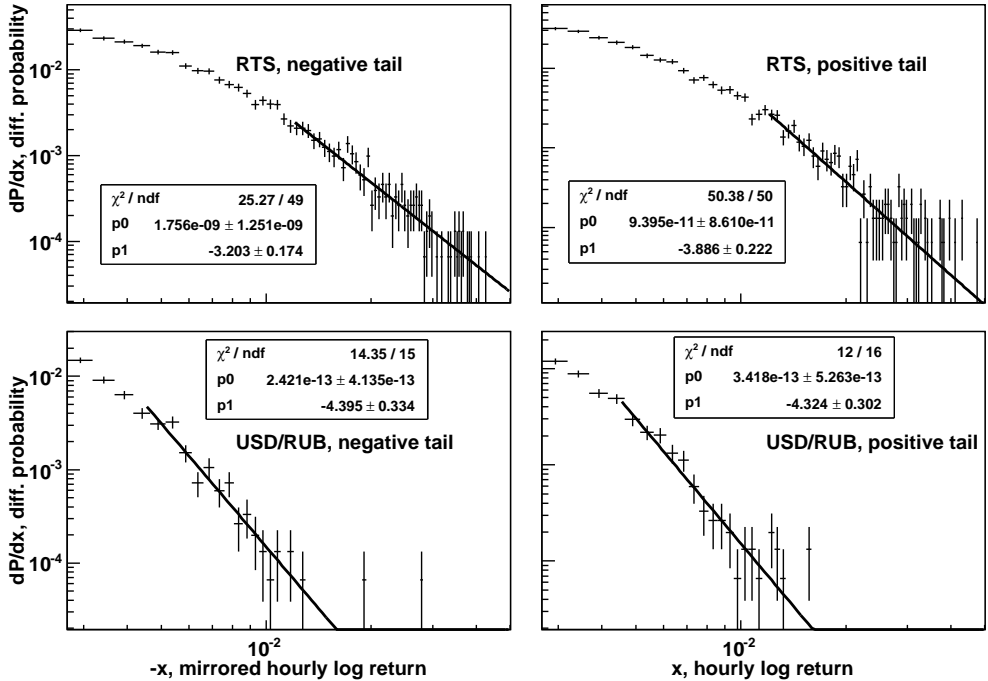}
\caption{Differential probability density distribution of hourly logarithmic returns. Top: futures on the RTS index, bottom: futures on the USD/RUB exchange rate. Left: tail of negative returns, mirror-reflected around vertical axis. Right: tail of positive returns. Results of least-square fits are for the power law parameterization, with $p_1$ being the power law exponent. Here and in all other figures, vertical bars show the magnitude of one standard deviation.
}
\label{fig:tails2x2}
\end{figure}

The requirement that all five markets have data for intermarket correlation analysis leaves us with the data set starting in February 2009. Unless specified otherwise explicitly, the data set under study is that from February 2009 through April 2013. The data set used in the analysis consists of 13.5-15 thousand hourly data points for each of the instruments. The immediate object of this correlation analysis is the time series of logarithmic returns on the hourly time scale, constructed using the hourly close of price $p(t)$:
\begin{equation}
x(t|\Delta t) = \ln(p(t)/p(t-\Delta t)) = \ln(p(t)) - \ln(p(t-\Delta t)),
\end{equation}
where $\Delta t$ is the time interval from the close of the previous hour having data to the given hour, normally one hour.

At the expiration of the futures contract, a switch is done to the next one which at that point has three more months (one month in case of Brent) 
till expiration.
Returns affected by the switch are always excluded from the analysis. 

For a particular hour to be included into the intermarket analysis, we require that all five markets of interest have data within that hour. If data from
at least one market are missing during a certain hour, the return data from that hour for all markets do not take part in the intermarket studies. 
This selection condition imposes a liquidity requirement on the data.

Fig. \ref{fig:tails2x2} presents data on the tail behavior of the differential probability density distribution of the logarithmic returns and
allows one to study tail asymptotics, to verify the convergence of quantities we are after.
Futures on the RTS index and the USD/RUB exchange rate are taken as typical examples.
The linearity of the tails on the log-log plot suggests a power law dependence.
Least-square fits with a power-law model
\begin{equation}
{\,dP}/{\,dx} = p_0 x^{p_1}
\end{equation}
yield parameters (shown in the insets) which indicate that one can safely rely on the convergence of first and second moments for these markets\footnote{The more negative is the power law exponent, the  thinner are the tails. For the $n$-th moment to exist, an integral $\int_{-\infty}^{+\infty} x^n p(x) \,dx$ has to converge, so if $p(x)$ follows a power law on the tails, $p(x) \propto x^{\alpha}$, $n+\alpha<-1$ ensures convergence.}, as $p_1<-3$.

The power exponents are within the expectations based on recent reviews \cite{powerLaw}. These exponents place an upper bound on the asymptotic complexity of statistical
analyses possible on the data, and consequently, on the asymptotic complexity of any algorithms aimed at detecting and exploiting statistically significant features.
The fit parameters in Fig. \ref{fig:tails2x2} indicate that in the case of USD/RUB, this upper bound is higher and genuine three-point analyses are not precluded by lack of convergence the way they are in the case of the RTS index.

The correlation function we use, 
\begin{equation}
c(t_d|x_1,x_2) = E_t[x_1(t+t_d)x_2(t)],
\label{eq:c}
\end{equation}
is such that a market trend (non-zero $E_t[x(t)]$) would be detected as a non-zero auto-correlation effect, in accord with our goals of detecting exploitable market dynamics. Here, $E_t$ is the time-averaging operator and 
$t_d = t_1-t_2$
is the time lag variable, the argument of the correlation function.

\begin{figure}[t]
\epsfxsize=13cm
\epsfbox{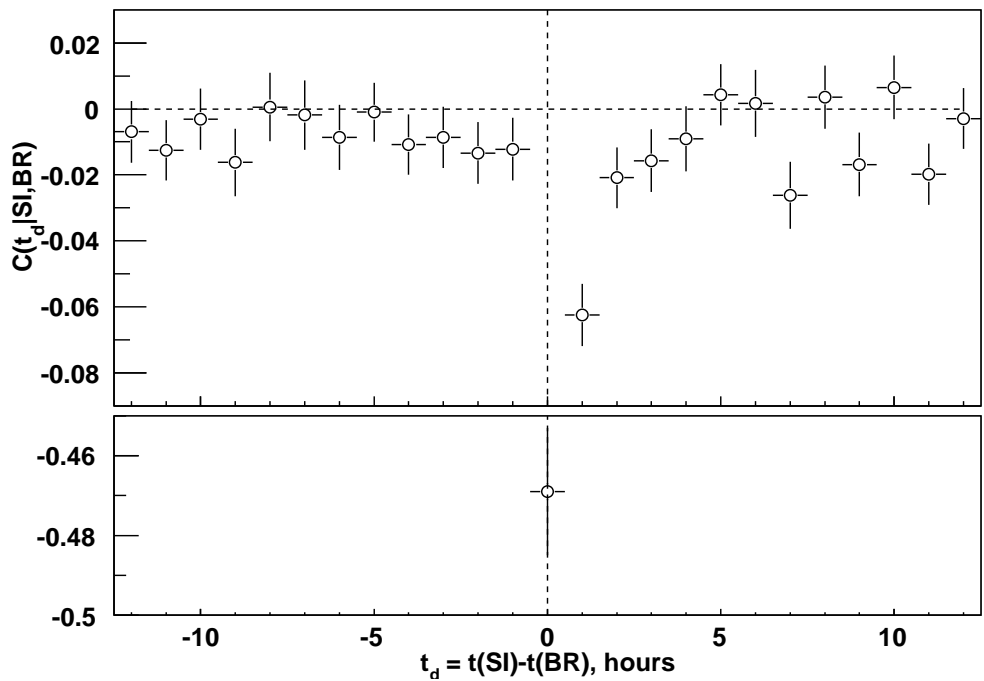}
\caption{Intermarket correlation function between nearest USD/RUB futures (SI) and nearest Brent futures (BR). Vertical bars are one standard deviation long.
}
\label{fig:SiBrCorrZoom}
\end{figure}

For intermarket studies, a normalized version of Eq.\ref{eq:c} is more convenient:
\begin{equation}
C(t_d|x_1,x_2) = c(t_d|x_1,x_2)/\sqrt{\mbox{Var}[x_1]\mbox{Var}[x_2]}.
\label{eq:C}
\end{equation}

An example of such a correlation function can be seen in Fig.\ref{fig:SiBrCorrZoom}.
The zero time lag value of this function is around -0.47, highlighting the role of the oil price 
as a positive driver for RUB value and therefore negative for USD/RUB, 
a macroeconomics feature familiar to all economic agents with interests in Russia.
We note that the asymmetry of the correlation with respect to a change in the sign of 
the time difference variable $t_d$, made statistically significant mostly by the presence of the non-zero 
correlation at the $t_d=1$  time lag.
This asymmetry indicates a leader-follower relationship between the instruments. 
Because $t_d$ is defined as $t(\mbox{SI})-t(\mbox{BR})$,
positive lags of one hour correspond to pairs of measurements where the SI datum is taken one hour after the BR. 
We interpret the lagged correlation to mean that the macroeconomically driven relationship between the instruments works out over a finite 
response time which is long enough to be visible with the data of hourly frequency. 
Fig.\ref{fig:SiBrCorrZoom} provides information to constrain models of such response.

\section{Results and discussion}

The statement that past rates of market return have no effect on future rates\cite{Fama} is known as weak form of the EMH. 
The leader-follower relationships, such as shown in Fig.\ref{fig:SiBrCorrZoom}, being a form of predictive inter-market dependency, falsify the weak form of the EMH and therefore, the stronger forms of it as well. Indeed, such effects can be exploited to shift the balance of profits and losses of informed market operators towards profit. 

Based on our experience of interacting, in the course of our consultancy business, with real economy decision makers exposed to FX risks,  we believe it would be untrue to say that all market participants are 
equally well informed about these effects. Nor are they equally well equipped to become informed. 
The EMH describes an interesting, but idealized condition, closely related to the concept of equilibrium. The subject of our study is not the validity of the EMH, but rather, quantifying the degree of its {\em invalidity} and presenting that as a potentially interesting addition to the familiar sets of econometric data.  

\subsection{Time-averaged picture of the leader-follower effects}
Fig.\ref{fig:SiBrCorrZoom} suggests that if one is interested in leader-follower effects on the hour time scale, then the correlation coefficient at a single-hour time lag might be the best single quantity to capture the effect. The rest of the correlation functions confirm this, therefore, to make presentation more economic, we will focus on this quantity.

\begin{table}
\label{tab:leader-follower}
\begin{tabular}{|c|c|ccccc|}
\hline
\multicolumn{2}{|c|}{}              & \multicolumn{5}{|c|}{follower} \\
\cline{3-7}
\multicolumn{2}{|c|}{}              & RTS & Brent & EUR/USD & EUR/RUB & USD/RUB \\
\hline
\multirow{2}{*}{leader} & RTS     &     &       &         &   -5.9  &   -4.3  \\
                        & Brent   &     &       &         &   -6.5  &   -5.6  \\
                        & EUR/USD &     &       &         &         &   -4.4   \\
                        & EUR/RUB &     &       &         &         &          \\
                        & USD/RUB &     &       &         &   4.2   &          \\
\hline
\end{tabular}
\caption{Strength and direction of the leader-follower effect in the hourly futures data under study. 
The number reported is the correlation strength at one hour lag, expressed in the units of standard deviation of the statistical precision of its measurement. Included here are all effects stronger than 3 standard deviations. The data are time-aggregated over the entire period of observation.}
\end{table}

Table 1 presents time-integrated data on the significance and direction of the leader-follower effect
for the period of observation. Only statistically significant effects are shown. 
None of the signals is in the diagonal of the  table i.e. in the autocorrelation. The effects are strictly asymmetric: a leader is never also a follower and vice versa. RTS and Brent both are leaders for EUR/RUB and USD/RUB, while USD/RUB is a leader to EUR/RUB and EUR/USD is a leader to USD/RUB. EUR/USD does not pass the signal strength threshold as a leader to EUR/RUB.

As to the sign of the effects reported in Table 1, they are all of the ``tail'' type, that is,
the sign of the correlation coefficient at the non-zero time lag is the same as the sign of the zero-lag correlation, so that 
the lagged correlation broadens the correlation peak.

If a market agent, already present in the market and conducting random activity, comes into possession of information represented by Fig.\ref{fig:SiBrCorrZoom} and
adjusts their operations accordingly without changing the overall volume, we assume that the hour-time-frame volatility (zero lag peak) will remain unchanged 
since the overall volume of operations has not changed.
The effect of that trader's operations will be to reduce the non-zero lag correlation coefficients, since one's buying pushes prices up and one's selling pushes them down, and thus one's actions to take advantage of an opportunity reduce the ``amount of opportunity'' available to all, the non-zero correlation at non-zero lag being a measure of that opportunity.

In the particular case of a hypothetic long RUB, long Brent portfolio -- a ``risk-on'' one -- the correlation features in Fig.\ref{fig:SiBrCorrZoom} would broaden
the auto-correlation function of the portfolio, compared to the EMH ideal,
 creating the phenomena known as market trends, e.g. creating predictability of future as a repetition of the past.
Since variance on a particular time scale is an integral of a correlation function with lags up to that time scale\cite{vanMarcke}, the narrowing of the correlation function
of the portfolio will result in a reduction of longer time scale variances (volatility).

Such arguments qualify speculation in markets with tail-type correlation as ``benign'' in a sense that it is possible, under such conditions,  to reduce predictability (increase ``efficiency'') 
while at the same time reducing
longer time scale volatility, as if speculators, to be profitable, were injecting stabilizing oscillation into the otherwise trending markets. Had the correlation shape been oscillating instead, the conclusion would have been the opposite, with speculators ``rocking the boat''.

\subsection{Time evolution of the leader-follower effects}
\begin{figure}[th]
\begin{tabular}{cc}
\epsfxsize=6.8cm
\epsfbox{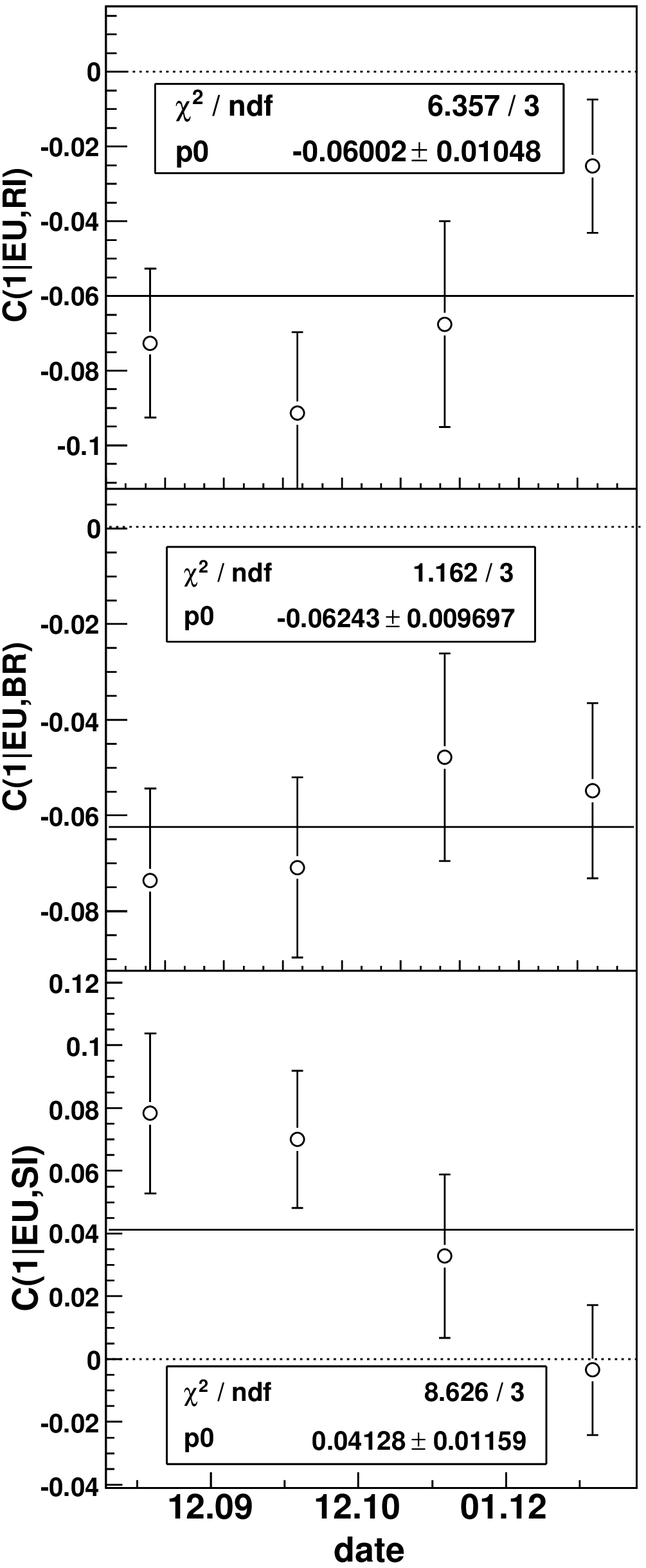} &
\epsfxsize=6.8cm
\epsfbox{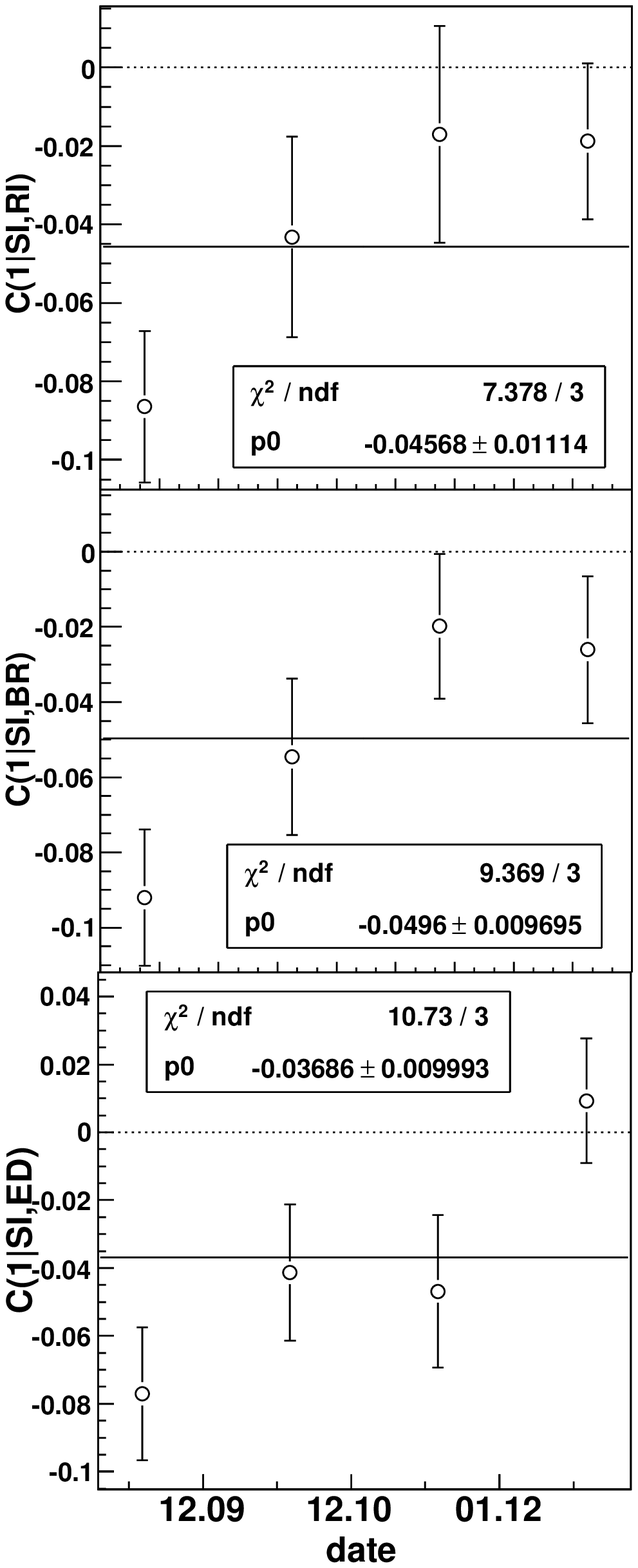} \\
\end{tabular}
\caption{History of  the 1-hour-lag correlation coefficient, reported to be significant in Table 
\ref{tab:leader-follower}. Left: correlation of EU, the EUR/RUB futures, with RTS, Brent, and USD/RUB. Right: correlation of SI, the USD/RUB futures,
with RTS, Brent, and EUR/USD.
Every data point represents a year. Solid horizontal lines show least-squares fits of the data sets with
constants ($p_0$), with $\chi^2$ per number of degrees of freedom shown in the plots. Vertical bars are one standard deviation long.
}
\label{fig:History}
\end{figure}

The natural next question is whether the effects reported in Table 1 are stable in time. 
To answer that, we have aggregated the data into four equal periods, each one year long, starting in February 2009 and ending in February 2013.
As in Table 1, we report one correlation coefficient at a time, the one at the time lag found by inspection of the time-integrated correlation
functions to carry the most significant signal. This is the single hour time lag. Fig.\ref{fig:History} presents results and statistical accuracy of
the measurement.

Upon visual inspection, five out of six time dependencies in Fig.\ref{fig:History} can be said to show a decrease in signal
strength over time. The only exception is correlation of EUR/RUB with Brent.

Goodness-of-fit or $p$-value is the probability of obtaining, even for a correct model, of a $\chi^2$ value that exceeds the value actually observed. This depends both on the $\chi^2$ and on the number of degrees of freedom.
With 3 degrees of freedom, for a fit as bad as  $\chi^2=6.4$,  top left plot in Fig.\ref{fig:History}, this is 0.094, whereas for $\chi^2=10.7$, bottom right, it is 0.013. All plots except for the one characterizing the correlation of EUR/RUB with Brent, middle left, make the temporal constancy of the effect an unlikely hypothesis, with goodness of the fit varying between the two $p$-values just quoted. 

In all six panels of Fig.\ref{fig:History}, whether or not the temporal non-constancy of data is judged as significant, the visual trend of the data points is invariably towards the dashed line representing the zero level of the effect, as the time goes on.

\subsection{Liquidity and the ordering of leader-follower pairs} 

\begin{figure}[th]
\epsfxsize=15cm
\epsfbox{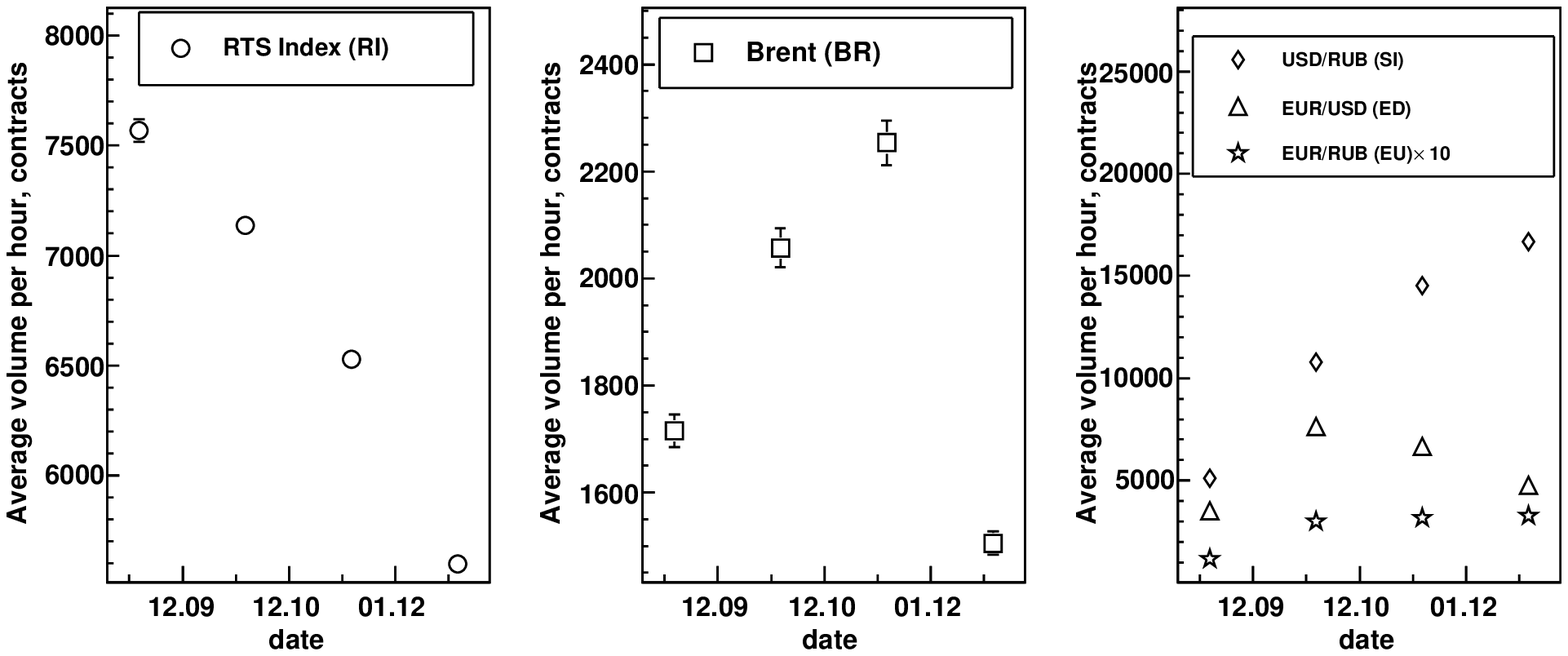} 
\caption{Volumes vs time. Hourly average number of contracts traded in a given year is shown as a function
of time.
}
\label{fig:Volumes}
\end{figure}

Liquidity for the five instruments under study is illustrated in Fig.\ref{fig:Volumes}.
Thousands of contracts per hour are traded for each instrument.

Does the volume have anything to do with who becomes the leader, and who -- the follower?
Not necessarily.
The most liquid instrument, USD/RUB, is a follower to RTS, Brent and EUR/USD. 
EUR/USD and Brent have volume outside FORTS, but RTS does not, and yet the RTS leads USD/RUB.
More liquid USD/RUB leads EUR/RUB.

The steady decline in the volumes of futures linked to the Russian stock market index, RTS, mirrors years-long 
decline of interest in Russia's stock market, allowing us to eliminate traditional asset managers from the list of 
``suspects'' responsible for the efficiency growth in FORTS.

Moreover, we hypothesize that the decline in the strength of leader-follower correlations involving RTS as a leader 
(Table 1) may be caused by
a peculiar decline in ``status'', previously enjoyed by the stock market, which may be losing ground in competition with  domestic real estate and RUB-denominated long-term bank deposits.

Along the same lines, one may suspect that crude oil and USD are similarly ``losing status'', relatively, of course 
-- if so, that can be
interpreted as a sign of growing confidence of Russian economic decision makers and growing complexity of the economy.

\subsection{Towards an inefficiency yardstick}

\begin{figure}[th]
\epsfxsize=15cm
\epsfbox{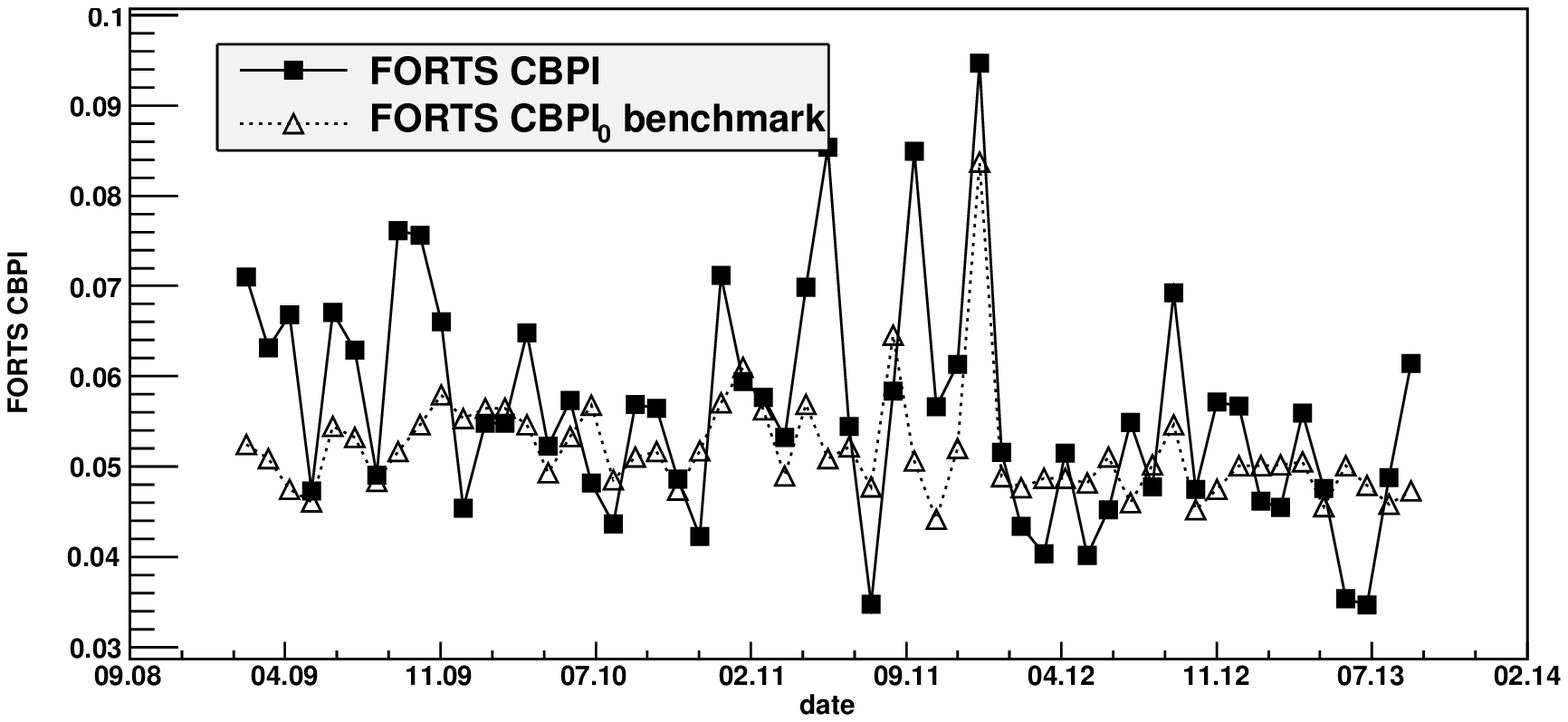} 
\caption{Monthly values of CBPI, the Correlation-Based Predictability Index for the FORTS markets under study, as a function of time, from 
February 2009 to August 2013.}
\label{fig:CBPI}
\end{figure}

Correlation-Based Predictability Index (CBPI) is a measure of market inefficiency with a specified time frame (time scale) and measurement interval. The measurement interval is the temporal coverage over which data are averaged to obtain a single CBPI quantity.  

CBPI is constructed from lagged autocorrelation and lagged intermarket correlation coefficients among the five futures contracts already introduced in section \ref{sec:data}. The lag equals one hour. The five time series form $5\times(5-1)/2 = 10$ unique intermarket pairs. However, due to permutations in the ordering of the pairs, the total number of intermarket combinations is 20. The permutations reveal new information since the correlation coefficient at positive time lag of one hour lag among time series A and B does not need to be the same as the negative one hour lag correlation coefficient of the same time series. To these 20 quantities, 5 one-hour lag autocorrelation coefficients are added to characterize self-predictability of the time series.

CBPI is defined as the average of the 25 above-described coefficients, each of which is taken by absolute value. Here is the general formula:  

\begin{equation}
\mbox{CBPI} = \sum_{i,j}|C(1|x_i,x_j)|/\sum_{i,j}1,
\end{equation}
where $C(1|x_i,x_j)$ is defined according to Eq.\ref{eq:C}.
To the extent efficient markets are understood as lacking memory, CBPI is a measure of market inefficiency in the selected group of markets on a particular time scale.

As any random quantity, CBPI is subject to fluctuations. Moreover, while growth in CBPI from zero measures the degree of predictability, a perfectly random market will usually have non-zero CBPI. Therefore, before CBPI can be used as a helpful yardstick of inefficiency, it too needs a benchmark.

As a non-negative quantity, CBPI is an average value statistic of an asymmetric distribution, describing an absolute value of the correlation coefficient corresponding to the time lag of one hour. The correlation coefficient is a random quantity which is equally likely to take both positive and negative values.
The distribution of its absolute value can be approximately described as the right half of a Gaussian distribution, centered at zero, whose $\sigma$ parameter equals the measurement accuracy for a typical one-hour lag correlation. 
Such a distribution has mean of $\sigma \sqrt{2/\pi}$. We take this quantity as a benchmark for CBPI and denote it 
$\mbox{CBPI}_0$. It models a would-be CBPI for the hypothetical markets where correlation coefficients are measurable with the same accuracy as in reality, but without any predictability effects. Naturally, comparison of CBPI with $\mbox{CBPI}_0$ is only meaningful for sufficiently large statistical samples of such quantities.

Figure \ref{fig:CBPI} shows the history of CBPI for the FORTS markets under study. It is complementary to Figure \ref{fig:History} in that, while the latter aggregates data over time, showing a detailed view of specific market combinations, the former aggregates data over markets (in a particular way), showing a detailed view of time evolution. 
The correlation effects already seen in Fig.\ref{fig:History} to be most pronounced in the first year of data, are also seen to leave a footprint in Figure \ref{fig:CBPI}.
 A study with a longer time span, covering 14 foreign exchange rate time series for the past 10 years, presented by one of us elsewhere\cite{ForexAutomaton}, paints a picture consistent with a long-term global decline of the inefficiency.

\section{Conclusions}
We observe statistical significant deviations of Russian futures markets from the conditions prescribed by the weak form of the Efficient Market Hypothesis. 
The strongest predictive inter-market correlations have been, on  the hourly time scale chosen,
 of the ``broad'' (tail-like) type, rather than of the oscillating type.
Under these conditions, any trading activity successful at  exploiting the predictive features (and thus, contributing to their reduction)
also reduces, everything else being equal, the longer-range volatility.

We  observe a statistically significant trend toward disappearance of the predictive leader-follower features in the course of the past four years. Only time will show whether this does not end up being a swing into the opposite extreme, namely that of the oscillatory predictability.

Alternatively, if we view the disappearance of leader-follower effects as a form of a {\em leadership crisis}, 
we may expect a return of the effects, perhaps with new leaders, in due course.

\section*{Acknowledgement}

This research has been supported by Maraging Partners LLC,  a Moscow-based consultancy which creates value for real economy businesses by developing efficient currency risk hedging strategies and by exploiting opportunities for excessive risk-adjusted returns, timing portfolio allocation on a quantitative, algorithmic and discretionary basis.
We are grateful to Felix Savchenkov for critical remarks which helped improve the presentation.

\end{document}